\documentclass[12pt,a4paper]{article}
\usepackage{amssymb,amsthm,cite,graphics}

\title{Feedback-induced stationary localized patterns in networks of diffusively coupled bistable elements} 

\author{Nikos E. Kouvaris and Alexander S. Mikhailov
\\
Department of Physical Chemistry, 
\\Fritz Haber Institute of the Max Planck Society
\\Faradayweg 4-6, D-14195 Berlin, Germany}
\date{\small nkoub@fhi-berlin.mpg.de}

\begin{document}
\maketitle
%
%
%
\begin{abstract}
Effects of feedbacks on self-organization phenomena in networks of diffusively coupled bistable elements are investigated. For regular trees, an approximate analytical theory for localized stationary patterns under application of global feedbacks is constructed. Using it, properties of such patterns in different parts of the parameter space are discussed. Numerical investigations are performed for large random Erd\"os-R\'enyi and scale-free networks. In both kinds of systems, localized stationary activation patterns have been observed. The active nodes in such a pattern form a subnetwork, whose size decreases as the feedback intensity is increased. For strong feedbacks, active subnetworks are organized as trees. Additionally, local feedbacks affecting only the nodes with high degrees ({\it i.e.} hubs) or the periphery nodes are considered.
\end{abstract}


%
%
%
%
\section{Introduction}
Nonequilibrium pattern formation has been extensively studied in distributed active media \cite{pismen-book-06,kapral-showolter-book-95,mik-book-I}. In excitable or bistable media, stationary localized patterns can develop. They represent domains with high activator density surrounded by large regions where the density of the activator is much lower. Such structures have been theoretically considered \cite{KOG80,ITO92} and experimentally observed in chemical reactions \cite{LI96,LEE94} and in semiconductors \cite{NIE92,TAR00}. They have also been studied under the global feedback control \cite{KRI94,MID92}. 
\par
Control of nonequilibrium patterns, as well as their purposeful design, is an essential issue that has attracted much attention \cite{schoell-handbook,MIK06,VAN08}. Global or local feedback control schemes serve as a standard method used for this purpose. Typically, global feedback requires a common control signal, generated by the entire system and applied back to all its elements. In contrast, local feedback control does not directly affect the entire system, but is applied only to some of its elements. Various feedback schemes have been used in theoretical studies \cite{KRI94,BAT97,BAT97a} and in the experiments \cite{KIM01,WANG01,SAK02,BER03}, either for stabilizing existing unstable patterns or for inducing new kinds of patterns that do not existing in the absence of feedback.
\par
Self-organization phenomena, such as epidemic spreading \cite{PAS01b,BAR04,COL07,COL08,barrat-book-08}, clustering \cite{NAK09} and synchronization \cite{ARE08} of oscillators, Turing patterns \cite{NAK10,Wolfrum2012a} or traveling and pinned fronts \cite{KOU12}, have been studied in reaction-diffusion systems organized in complex networks. Moreover, some effects of control by global feedbacks have been previously investigated for networks systems. It has been demonstrated, for example, that turbulence in oscillator networks can be suppressed \cite{GIL09a} and hysteresis of Turing network patterns can be prevented \cite{HAT12} when such feedbacks are applied. 
\par
As we show in this Letter, feedback control may also suppress spreading of activation in networks of bistable elements, and leads to the emergence of localized stationary patterns which resemble stationary spots in continuous media. The approximate analytical theory of such phenomena could be constructed for regular trees and systematic numerical simulations for random Erd\"os-R\'enyi (ER) and scale-free (SF) networks were undertaken. Both global and local feedbacks, applied to a subset of network nodes, were considered. Properties of developing stationary patterns, depending on control conditions, have been explored. 
%
%
%
%
\section{Bistable systems on networks}
Classical continuous one-component bistable media are described by $\dot{u}(\mathbf{x},t) = f(u,h) + D \nabla^{2} u(\mathbf{x},t)\,,$ where $u(\mathbf{x},t)$ is the activator density, $D$ is the activation diffusion constant and function $f(u,h)$ specifies the local bistable dynamics. For example, this function can be chosen as $f(u,h)=u(h-u)(u-1)\,$, where the parameter $h$ determines the activation threshold and plays an important role in front propagation. 
\par
In network-organized systems, the activator species occupies the nodes of a network and can be diffusively transported over network links to other nodes. The architecture of a network is described in terms of its adjacency matrix $\mathbf{T}$, whose elements are $T_{ij}=1$, if there is a link connecting the nodes $i$ and $j$ ($i,j=1,...,N$), and $T_{ij}=0$ otherwise. We will consider processes in non-directed networks, where the adjacency matrix is symmetric ($T_{ij} = T_{ji}$). Generally, the network analog of one-component bistable system is given by
\begin{equation}\label{eq:rdnetT}
\dot{u}_i  = f(u_i,h_i) + D \sum_{j=1}^{N}\!\left(T_{ij}u_j - T_{ji}u_i \right)\,,
\end{equation}
\noindent where $u_i$ is the amount of activator in network node $i$ and $f(u_i,h_i)$ describes the local bistable dynamics of the activator. The last term in eq.~(\ref{eq:rdnetT}) takes into account diffusive coupling between the nodes and the coefficient $D$ characterizes the rate of diffusive transport of the activator over the network links. An essential property of a node $i$ is its degree $k_{i}$ (the number of connections) given by $k_i=\sum_jT_{ji}$. 
 \par
We introduce the global feedback through the parameter $h$, {\it i.e.} as
\begin{equation}\label{eq:feed1}
h = h_0 + \mu(S-S_0)\,,
\end{equation}
\noindent where $\mu>0$ is the intensity of the feedback, $S=\sum_{j=1}^{N}\!u_j\,$ is the total activation of the network and $S_0$ is a parameter defining the size of localized patterns. In our simulations, it was taken equal to the number of the nodes which were initially activated. Hence, the threshold $h$ depends now on the total activation. It increases when more nodes are activated, so that a negative feedback is realized. 
%
%
%
%
\section{Regular tree networks}
We first investigate the system (\ref{eq:rdnetT}) without feedback control, where spreading or retreating activation fronts can be found. Fronts can also get pinned, thus forming stationary patterns. Previously, we have derived the pinning condition for regular trees depending on the parameters $k$ and $D$ \cite{KOU12}, but now we would also need to know how the pinning behavior depends on the threshold $h$. We analyze this behavior and determine how it is affected by the feedback.
\par
Let us consider the system (\ref{eq:rdnetT}) on a regular tree with the branching factor $k-1$. In such a tree, all nodes, lying at the same distance $l$ from the origin, can be grouped into a single shell. Suppose that we have taken a node which belongs to the shell $l$. This node should be diffusively coupled to $k-1$ nodes in the next shell $l+1$ and to just one node in the previous shell $l-1$, as described by the equation
\begin{equation}
\dot{u}_{l} = f(u_{l},h_{l}) + D (u_{l-1}-u_{l}) + D(k-1)(u_{l+1}-u_{l})\,,
\label{eq:kchain}
\end{equation}
\noindent where $u_l$ is the density of the activator in the shell $l$. Note that for $k=2$, eq.~(\ref{eq:kchain}) corresponds to a one-dimensional chain of coupled bistable elements, where pinned fronts have been previously investigated \cite{ERN93,MIT98}. However, the respective approximate analytical theory for the trees ($k>2$) has been only recently developed \cite{KOU12}. 
\begin{figure}[t]
\includegraphics{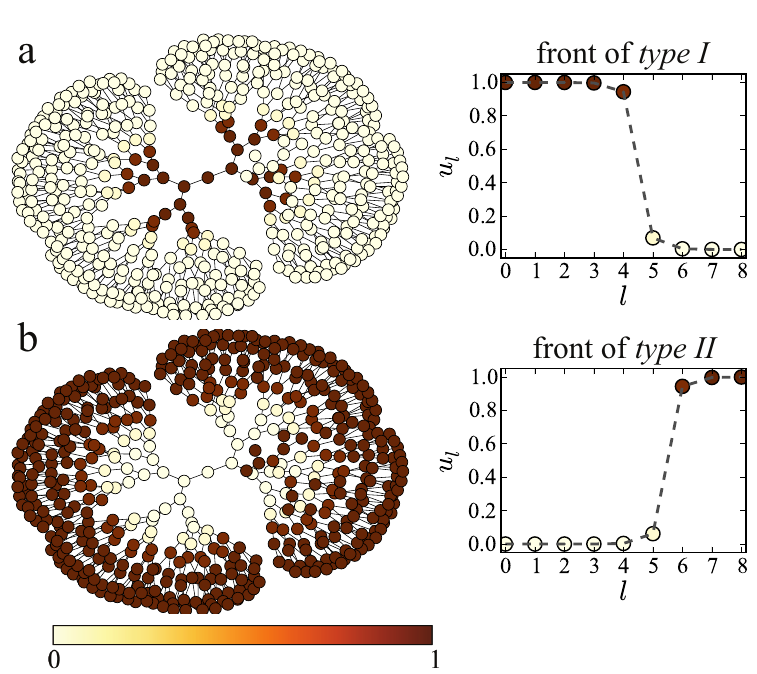}
\caption{(Color online) Stationary patterns corresponding to pinned fronts of {\it type I} (a) and {\it type II} (b) in a regular tree with the branching factor $k-1=2$. By grouping the nodes with the same distance from the root into a single shell, the two types of fronts can be displayed as shown on the right side. The parameters are (a) $h=0.35$ and (b) $h=0.65$; $D=0.02$}
\label{fig:trees}
\end{figure}  
\begin{figure}[b!]
\centering
\includegraphics{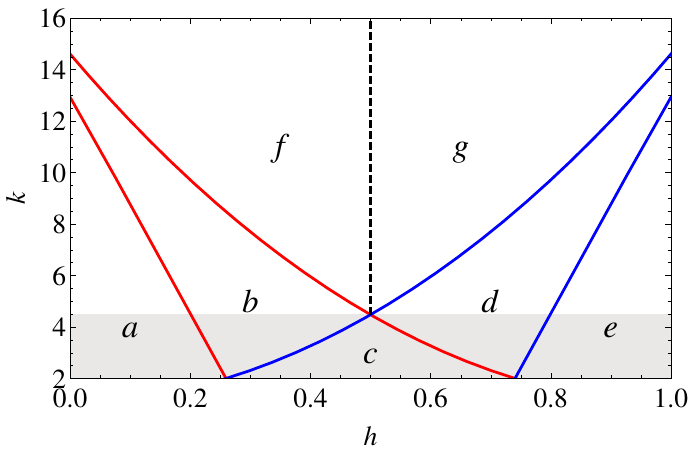}
\caption{(Color online) Bifurcation diagram in the plane ($h,k$) for $D=0.02$. Red curves correspond to eqs.~(\ref{eq:sn1}), while blue curves correspond to eqs.~(\ref{eq:sn2}). The black dashed line determines the boundary on which the velocities of both types of fronts moving towards the tree root become equal.}
\label{fig:bifs}
\end{figure}  
\begin{figure}[t]
\centering
\includegraphics{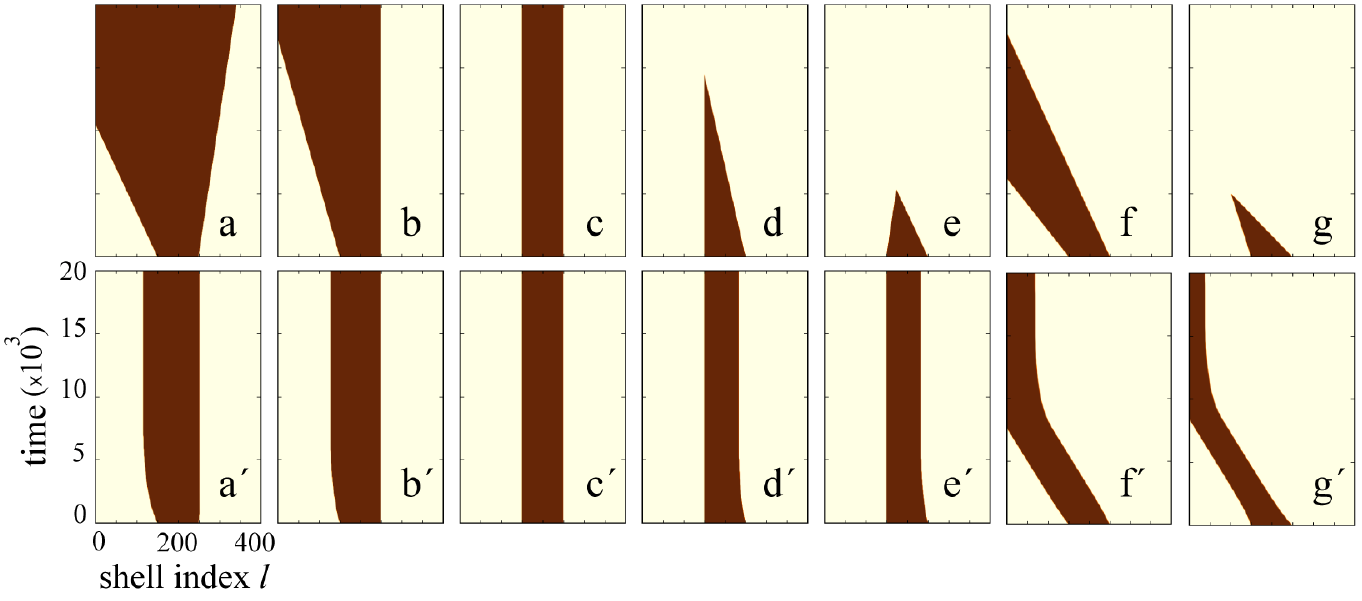}
\caption{(Color online) Evolution plots in absence (a--g) and in presence (a$^\prime$--g$^\prime$) of global feedback are shown for different regions of the bifurcation diagram (fig.~\ref{fig:bifs}). Global negative feedback had the intensity $\mu = 0.001$ and was characterized by $S_0 = 100$. The diffusion constant was $D = 0.02$. The degrees were (a--e) $k = 4$ and (f,g) $k = 13$. The thresholds were chosen as (a) $h=0.1$, (b) $h=0.25$, (c) $h=0.47$, (d) $h=0.7$, (e) $h=0.9$, (f) $h=0.4$ and (g) $h=0.7$. The same color coding as in fig.~\ref{fig:trees} is used.}
\label{fig:spacetime}
\end{figure}
\par 
Generally, both the activation front propagating from the root to the periphery of a tree ({\it type I}) and in the opposite direction, {\it i.e.} towards the tree root ({\it type II}), are possible. First, we derive approximate pinning conditions for these two types of fronts, depending on the branching factor $k-1$ and the parameter $h$. If diffusion is weak enough ({\it cf.} \cite{MIT98,KOU12}), a pinned front can be found by setting $\dot{u}_{l}=0$ in eq.~(\ref{eq:kchain}), so that we get $f(u_{l},h_{l}) + D(u_{l-1}-u_{l}) + D(k-1)(u_{l+1}-u_{l}) = 0\,$. Suppose that a front of {\it type I} is pinned and located at the shell $l=m$. The front is so sharp that the nodes in the lower $l<m$ shells are all approximately in the active state. Then, the activation level $u$ at the shell $m$ should approximately satisfy the condition $g(u) = f(u,h) + D ( 1 - ku ) = 0\,$. Stationary localized patterns are found within the parameter region where function $g(u)$ has three real roots. The boundaries of the pinning 
region in the plane $(k,h)$ are determined by the equations (see \cite{KOU12})
\begin{eqnarray}\label{eq:sn1}
k &=& \frac{u^4 + 2 u \left(D-u^2\right) + \left(u^2-D\right) } { D u^2 }\,, \nonumber\\
h &=& \left(\frac{D}{u^2}-1\right) + 2 u\,.
\end{eqnarray}
\noindent Thus, the fronts of {\it type I} are pinned within the region which lies between two red curves in the bifurcation diagram in fig.~\ref{fig:bifs} and comprises parts {\it b} and {\it c}. Stationary patterns around the tree root, corresponding to the pinned front of {\it type I} are shown in fig.~\ref{fig:trees}(a).
\par
Repeating the analysis for a front of {\it type II}, one can derive the pinning condition given in the parametric form by 
\begin{eqnarray}\label{eq:sn2}
k &=& \frac{u^4 + \left(D+u^2\right) - 2 u \left(D+u^2\right) } { D \left(1-u\right)^2 } \,,\nonumber\\
h &=& -\frac{D}{\left(1-u\right)^2} + 2 u\,.
\end{eqnarray}
\noindent The fronts of {\it type II} are pinned between two blue curves in fig.~\ref{fig:bifs}, {\it i.e.} in the areas {\it c} and {\it d}; an example of a stationary pattern in this case is shown in fig.~\ref{fig:trees}(b). 
\par
Thus, our approximate theory has allowed us to identify regions in the parameter plane ($h,k$) where stationary patterns are localized around the root or at the periphery of the trees, without the feedback control. Furthermore, using fig.~\ref{fig:bifs} we can consider evolution of various perturbations in different regions of the bifurcation diagram.
\par
The evolution plots in figs.~\ref{fig:spacetime}(a)--(g) show the development of the initial perturbation, localized in the shells between $l_{1}=150$ and $l_{2}=250$, in different regions of the bifurcation diagram in fig.~\ref{fig:bifs}. In absence of feedback control, for parameters within the region {\it a}, the activation spreads in both directions, towards the root and the periphery, eventually transferring the entire tree into the active state (fig.~\ref{fig:spacetime}(a)). For the parameters within the regions {\it b} and {\it c} of the bifurcation diagram stationary patterns are formed. In region {\it b}, the activation spreads only towards the root; once the root gets activated the pattern becomes stationary and includes all nodes in the shells $l_{2}\geq l\geq 0$  (fig.~\ref{fig:spacetime}(b)). In region {\it c}, the stationary pattern remains localized within the initial perturbation interval $l_{2}\geq l\geq l_{1}$ (fig.~\ref{fig:spacetime}(c)). In region {\it d}, the activation gets annihilated,
 because it is pinned on the root's side while retreating from the periphery (fig.~\ref{fig:spacetime}(d)). In region {\it e}, the perturbation retreats on both sides and disappears (fig.~\ref{fig:spacetime}(e)). In region {\it f}, the active domain is gradually broadened while traveling in the root direction, but it is also retreated from the periphery and finally vanishes. In region {\it g}, the local activation is shrinking while traveling in the same direction. 
\par
Thus, in absence of feedback control, local perturbation gives rise to stationary patterns only for the parameters within the regions {\it b} and {\it c} of the bifurcation diagram in fig.~\ref{fig:bifs}. As we show below, in presence of global feedback, the threshold $h$ is not constant but varies with the total activation, according to eq.~\ref{eq:feed1}. Thus, different regions of the bifurcation diagram can be transversed, until a localized stationary pattern is established.
%
%
%
%
\begin{figure}[t!]
\centering
\includegraphics{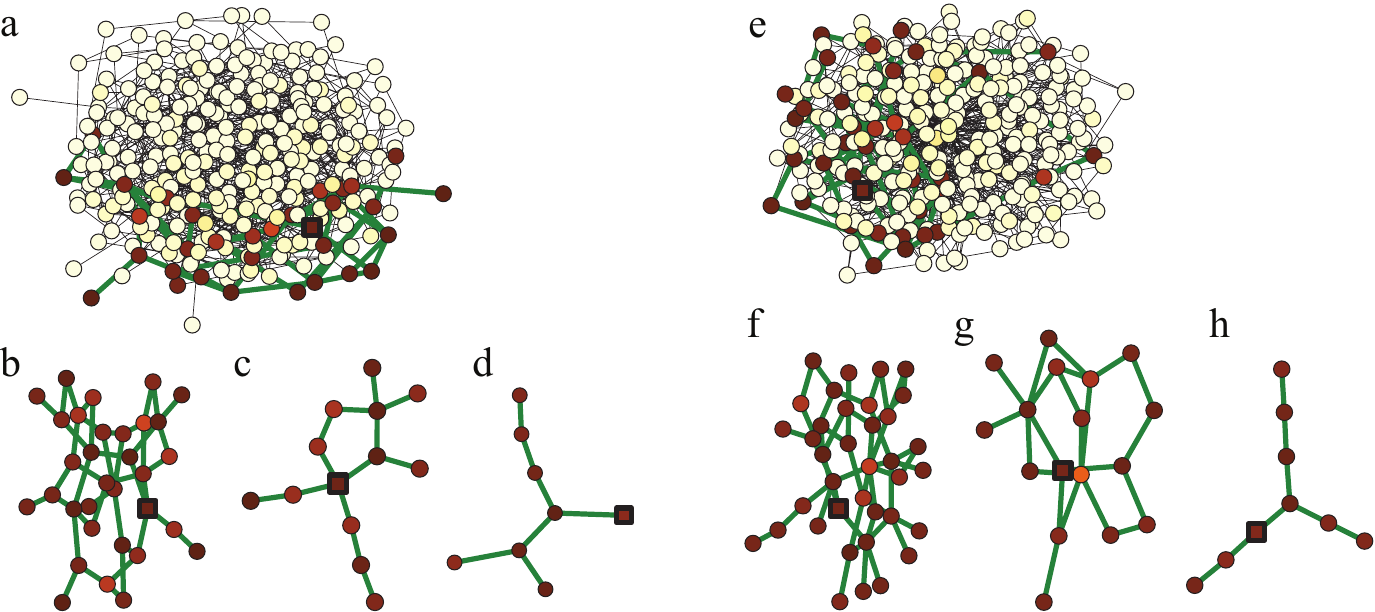}
\caption{(Color online) Localized stationary patterns in an ER network (a--d) and in a SF network (e--h) with $\langle k\rangle=6$ and $N=300$. The black square indicates the node at which the activation was initially applied. Activation patterns of entire networks are shown for $\mu=0.004$ in the upper row (a,e). The bottom row shows activated subnetworks at the increasing feedback intensities $\mu=0.004$ (b,f), $\mu=0.008$ (c,g) and $\mu=0.012$ (d,h). Other parameters are $h_0=0.1$ and $D=0.02$; the same color coding as in fig.~\ref{fig:trees} is used.}
\label{fig:randnets}
\end{figure}
\par
When global feedback (eq.~(\ref{eq:feed1})) is applied, stationary patterns develop in all regions of the bifurcation diagram. For the degrees $k$, below the intersection of curves (\ref{eq:sn1}) and (\ref{eq:sn2}) at $h=0.5$ (gray shaded region in fig.~\ref{fig:bifs}), such patterns stay localized around the initial perturbation. When the parameter $h_0$ is chosen within region {\it a}, the activation starts to spread in both directions. As a result, the threshold $h$ increases, thus transferring the dynamics into the region {\it b}, where the outer front gets pinned and spreading continues only inwards to the root. As $h$ increases further, the region {\it c} is entered, where the pattern becomes stationary (fig.~\ref{fig:spacetime}(a$^\prime$)). In fig.~\ref{fig:spacetime}(b$^\prime$), a similar scenario takes place, with the feedback gradually shifting the dynamics from region {\it b} to {\it c}. In region {\it c}, the initial activation remains frozen (figs.~\ref{fig:spacetime}(c$^\prime$)). In region {\
it d}, the activation does not spread in the direction of the root, but it is retreated from the periphery. Therefore, $h$ decreases and a stationary pattern is formed, once $h$ enters region {\it c} (fig.~\ref{fig:spacetime}(d$^\prime$)). Choosing $h_{0}$ within region {\it e}, we find that (fig.~\ref{fig:spacetime}(e$^\prime$)) the activation first shrinks on both sides. Then, it gets pinned on the root's side and after that the same evolution as in fig.~\ref{fig:spacetime}(d$^\prime$) takes place.
\par
In the trees with the larger branching factor, {\it i.e.} the larger degrees $k$, feedback induces localized patterns which travel towards the root. Once the root becomes activated, the size of the patterns changes in such a way that the threshold $h$ enters region {\it b} of the bifurcation diagram and the patterns become stationary, as shown in figs.~\ref{fig:spacetime}(f$^\prime$) and (g$^\prime$). This behavior, leading to pattern formation around the tree root, is also observed for the larger degrees $k$, where pinning of fronts cannot occur. 
\par
Summarizing, we have shown that global negative feedback can induce stationary localized patterns in the trees, even when they do not exist in its absence. The pattern formation mechanism could be understood by using the bifurcation diagram in fig.~\ref{fig:bifs}. 
%
%
%
%
\section{Random networks}
Effects of global negative feedback have been numerically studied for random ER and SF networks (fig.~\ref{fig:randnets}). As we have seen, the activation that was applied to one node (marked by a square in fig.~\ref{fig:randnets}) started to spread out. This produced a growing cluster of activated nodes. Its growth was however accompanied by an increase of the negative feedback. Consequently, the threshold $h$ increased with the size of the cluster, making the activation more difficult. As a result, the growth of the activated cluster was slowed down and finally stopped. Thus, a stationary activation pattern, representing a small subnetwork embedded in the entire network (see fig.~\ref{fig:randnets}) became formed.
\par
By retaining only the nodes with sufficiently high activation level $u>0.7$, active subnetworks can be identified. A sequence of such subnetworks, obtained under an increase of the global feedback intensity, is shown in figs.~\ref{fig:randnets}(b--d) for an ER network and in figs.~\ref{fig:randnets}(f--h) for a SF network. As the feedback gets stronger, active subnetworks decrease in size and approach a tree structure. Our statistical analysis has revealed that, in both ER and SF networks, the average size of active subnetworks approximately followed the power law $S\propto\mu^{-1}$. 
%
%
%
%
\subsection{Local feedback schemes}
\begin{figure}[t!]
\centering
\includegraphics{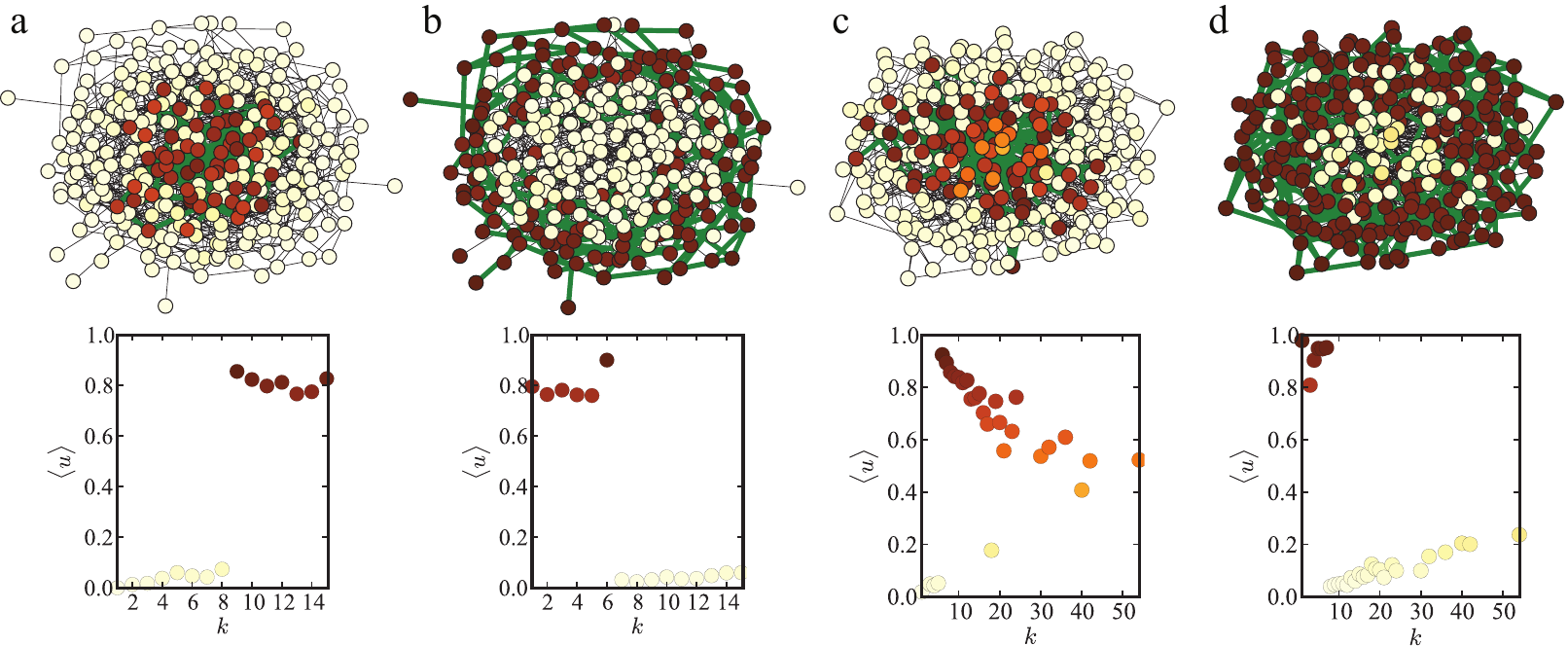}
\caption{(Color online) Stationary patterns localized on the hubs (a,c) and on the periphery nodes (b,d) of the ER (a,b) and the SF (c,d) networks with $\langle k\rangle=6$ and $N=300$ for $\mu=0.01$. Dependences of the mean activation $<u>$ on the degree $k$ of the nodes in the localized patterns for the feedback applied only to the periphery nodes with $k<k_0$ where  (a) $k_{0} = 9$ and (c) $k_{0} = 6$, or for the feedback applied to the hubs with the degrees exceeding (b) $k_{0} = 6$ and (d) $k_{0} = 7$. Other parameters are $h_0=0.1$ and $D=0.02$; the same color coding as in fig.~\ref{fig:trees} is used.}
\label{fig:randnets2}
\end{figure}
So far, we have assumed that the negative feedback was global. However, it is also possible that the feedback acts only on a subset of nodes. Suppose for example that the parameters $h_i$, which control the dynamics of individual nodes $i$, have different sensitivity to the global control signal depending on the degrees of such nodes, {\it i.e.} that
\begin{equation}\label{eq:feed2}
h_i = h_0 + \mu B(k_i)(S-S_0)\,,
\end{equation}
\noindent where $B(k_i)$ is the step function, $B(k)=1$ if $k<k_{0}$ and $B(k)=0$ otherwise. In this case, the negative feedback is applied only to the periphery nodes with small degrees $k$, where $k<k_0$. By introducing such local feedback, we can control directly the dynamics of periphery nodes, whereas hubs remain non-affected and have a constant threshold $h=h_0$ sufficient for activation spreading. The simulations show that, in this situation, the activation applied to a hub node starts to spread forming a cluster of activated nodes with high degrees $k\geq k_0$, whose further growth to the periphery nodes is suppressed by the feedback. Thus, an active stationary cluster, localized on the hubs and shown for an ER network in fig.~\ref{fig:randnets2}(a), becomes formed.
\par
Alternatively, we can choose function $B(k_i)$ as $B(k)=1$ for $k>k_{0}$ and $B(k)=0$ otherwise, so that the feedback is applied only to the hubs. Then, the evolution leads to the formation of a stationary pattern, localized on the periphery nodes, as shown for the ER network in fig.~\ref{fig:randnets2}(b).
\par
Similar behavior was observed when local feedbacks were applied to SF networks. Figure~\ref{fig:randnets2}(c) shows a stationary pattern, localized at the hubs of a SF network, with the negative feedback directly affecting the nodes with degrees smaller than $k_0=6$. However, if the feeback is applied to the hubs with the degrees higher then $k_0=7$, the active pattern is localized at the periphery nodes, as shown in fig.~\ref{fig:randnets2}(d).
%
%
%
%
\section{Discussion}
Our study has shown that control and purposeful design of nonequilibrium patterns is possible on the networks of bistable elements. Stationary localized patterns were found in the trees as well as in the random ER and SF networks, by applying different schemes of negative feedback. They consisted of nodes with high activator density, surrounded by nodes where the density of activator was much lower, and were representing the network analogs of stationary spots in classical reaction-diffusion media. Their size and structure could be controlled by varying the feedback intensity. 
\par
In the special case of regular trees, an analytical theory could be constructed, providing the pinning conditions as a function of control parameter. As revealed through numerical simulations, the global negative feedback can drive the system dynamics into the pinning regions, thus giving rise to the formation of stationary patterns.
\par
For random ER and SF networks, feedback-induced stationary patterns are localized on subnetworks of the entire system. The structure and the size of such subnetworks can be controlled by the feedback intensity. Their size decreases as the feedback becomes stronger and follows a power law. For sufficiently high feedback intensities $\mu$, the activated subnetwork approaches a tree.
\par
For large random networks, we have also studied the effects of negative feedbacks which were local, {\it i.e.} acting on the subsets of nodes with small or large degrees. These feedback schemes resulted in stationary patterns which were not small subnetworks and consisted of large groups of  hub or periphery nodes.  
\par
Thus, the effects of different feedback schemes on dynamics of networks formed by diffusively coupled bistable elements were investigated. In the future, it would be interesting to extend this analysis and consider time-delayed feedback schemes, which may lead to oscillating patterns or other complex dynamical structures on the networks.

\section*{Acknowledgments} 
The authors would like to thank Prof. Hiroshi Kori for stimulating discussions. Financial support from the DFG Collaborative Research Center SFB910 ``Control of Self-Organizing Nonlinear Systems'' and from the Volkswagen Foundation in Germany is gratefully acknowledged.

%
%
%
%

\end{document}